\documentstyle[editedvolume]{crckapb} 

\begin{opening}
\title{Where Are We and Where Should We Go?}

\subtitle{Summary Reports of the Girona Conference Workshops}

\author{BUELL  T. JANNUZI}
\institute{National Optical Astronomy Observatories\\
           P.O. Box 26732, Tucson, AZ 85726, USA}
\author{Alan P. Marscher}
\institute{Department of Astronomy, Boston University\\
725 Commonwealth Ave., Boston, MA 02215, USA}
\author{Martin Pohl}
\institute{Max-Planck-Institut f\"ur Extraterrestrische Physik \\
Postfach 1603, 85740 Garching, Germany} 
\author{Alex G. Smith}
\institute{University of  Florida 
Department of Astronomy, 211 Space Sciences Building, Gainsville, FL
32611, USA}

\end{opening}

\runningtitle{Workshop Summaries}

\begin{document}



\section{Introduction}

After nearly four days of listening to our colleagues describe their
latest results on jets, black holes, and blazars, the conference
organizers gave the participants a chance to take stock of how much
(or how little) has been learned about blazars and speculate about
future directions of research.  Four parallel workshops were organized
and conference attendees were free to choose at which session they
would like to present and debate their ideas for the duration of an
afternoon, with opportunities for coffee and cookies to stimulate the
discussion.  The chosen topics were ``Jet Models'', ``Multi-Frequency
Support for High-Energy Observations'', ``Monitoring of Blazars'', and
``Host Galaxies and Environments''.  While the individual sessions
focused on their assigned topic, the goals of all the sessions were to
identify points of agreement in the field and to set a course for
future research.  Following several hours of lively discussions, each
session chair presented to the entire conference a summary of their
own group's discussion.  In this collective summary we have tried to
reproduce the main opinions and ideas that were generated during the
workshops.  The individual workshop summaries are presented in the
following subsections and were prepared by the respective session
chairs: ``Jet Models'', Marscher; ``Multi-Frequency Support'', Pohl;
``Monitoring of Blazars'', Smith; and ``Host Galaxies and
Environments'', Jannuzi.

\section{Jet Models}

\subsection{Introduction}

What do you get when you lock up about twenty blazar theorists in a
room for an hour and a half? The answer at the outset was not clear:
100\% annihilation of theorists and their corresponding
anti-theorists? Anarchy followed by one strong-armed astrophysicist
taking over as dictator and forcing his/her ideas on everyone else?
General agreement on the status of the field and the things to do
next, with everyone leaving the room arm-in-arm, singing ``Give Peace
a Chance'' in perfect five-part harmony? I was very pleasantly
surprised --- shocked, even, but perhaps that's because I like models
with shocks so much --- that the result was much closer to the last of
these choices than the former two. It seems that there is a widespread
consensus on the next steps to take in jet modeling and simulation.
In the following brief report, I summarize the current status of the
field and the direction of future efforts that the participants in the
workshop feel would be most fruitful toward understanding blazars,
which we all know are the most interesting objects in the universe.
 
\subsection{Nature of the Jets}

While most models are based on the assumption that the jet is a plasma
flowing at a relativistic speed that is the same speed needed to
explain the observed apparent superluminal motion, this is not
necessarily true. In fact, there are some disturbingly rapid
variations (e.g., the recent factor-of-20 TeV flare in Mkn 421 on a
timescale of less than half an hour) that are pushing us toward ever
higher Doppler factors. One suggestion is that the underlying
phenomenon is an invisible particle beam with very high Lorentz factor
($10^4$ to $10^6$) that interacts somewhat with the external medium to
produce secondary, but visible features that move at the much slower
(Lorentz factor $\sim$ 3--20) speeds seen in VLBI images. It would be
natural for such beams to be composed of electrons and positrons, in
which case the circular polarization would be zero. Highly sensitive
observations of circular polarization could confirm or rule out an
e$^+$-e$^-$ jet.

\subsection{Jet Formation and Collimation}

Some progress --- much reported at the meeting --- has been made in
understanding the formation and collimation of jets. These usually
involve magnetohydrodynamic (MHD) calculations with varying
idealizations. In general, the more closely the jet is linked to the
accretion disk, the more ideal the calculations need to be in order to
make the problem tractable. Currently, the calculations use the
self-similarity approximation, and in the relativistic case the
magnetic field is force-free. The early results suggest that thin
accretion disks might not be able to drive powerful jets. Another
result is that radiating shocks are unlikely to form inside the fast
magnetosonic radius, at $\sim 10^{15}$--$10^{16}$ cm from the central
engine for a high luminosity jet.

Progress in this area will require first full two-dimensional, and
eventually 3-D, MHD simulations. Resistivity should be allowed so that
particles can cross field lines. What need to be studied are the
relationships between the jet cross-sectional radius, the jet power,
and the magnetic field on the structure of the jet.  In addition, it
is important to study the effects of radiative losses on the
dynamics. For electron-proton jets, this will require some knowledge
of the relevant plasma physics to determine the extent to which the
protons share their energy with the electrons.

\subsection{Simulations of Parsec- and Kiloparsec-Scale Jets}

At long last, there are now several groups carrying out time-dependent
relativistic hydrodynamical simulations of jets in blazars. Despite
some idealizations, these are already producing results that compare
favorably with the observations, as reported elsewhere in these
proceedings. The full range of parameter space needs to be explored in
order to study the effects of different forms of the external pressure
gradient, the reaction of the jet to different types of temporary
perturbations, and the dependence of jet appearance on the values of
the important physical parameters.

At kiloparsec scales, it appears that complete understanding of the
details seen on VLA images will require three-dimensional,
relativistic MHD simulations at high spatial resolution. This is a
rather ambitious goal, but seems now to be in reach.

\subsection{High-Energy Emission Models}

There is a lot of current activity among theorists trying to explain
the gamma-ray emission from blazars. In general, steady state
calculations are done very well, but computational difficulties have
impeded progress in the crucial time-dependent case that is so
important for comparison with observations. Realistic time-dependent
models need to include all the various radiative processes and sources
of seed photons for Compton scattering, the dynamics of jets and
shocks inside them, and the evolution of the energy distribution of
relativistic electrons. A number of groups have begun studies along
these lines.

\subsection{Jet Models: Conclusions}

It was striking to me that the theorists at the meeting had such a
clear, critical view of the current state of the field, as well as an
equally clear vision on what improvements to the models are likely to
produce more realistic versions.  Nevertheless, observers should
always understand that the theorists world is, by necessity, an ideal
one: the number of free parameters must be limited in order for
success in application to the observations to be considered credible.
The models can therefore never be expected to explain ``that little
wiggle'' in the light curve or ``that strange-looking elongated
feature'' in the radio image.  What the theorist hopes for is an
understanding of the basic physical processes that operate in an ideal
jet, and that the term ``ideal'' becomes less disconnected from
reality as the model is developed. Any observer who wants a closer
agreement with reality should heed the following advice: {\it Smooth
your data!}

In this vein, I wrote a limerick, the verses of which I sang
(croaked?) at the meeting. It is best heard (or, in this case, read)
with one's favorite bottle of spirits. I have added a chorus in
Spanish, sung to a well-known Spanish drinking song that should be
obvious.

\vskip 15mm

\centerline{\it The Theorist's Lament}
\vskip 3mm\noindent
There was a young theorist from Spain\\
Whose success with HD models was plain\\
But other theorists agreed\\
It wasn't MHD\\
So they ignored all the data it explained!\\
\\
Chorus:\\
Ay, ay, ay, ay\\
La vida de un te\'orico es desesperante\\
Crea modelos despanpanantes\\
Que luego son destruidos\\
Por datos extravagantes!\\
\\
There was this shock model of mine\\
That seemed to fit light curves just fine\\
But certain observers said\\
``The shock model is dead\\
'cause it can't fit this flux from '89''\\
\\[0mm]
[Chorus]
\vskip 5mm
The moral of this sad story is: Be kind to your neighborhood theorist, who
knows full well how idealized his/her models are and how poorly they
often fit the data. Besides, without the theorists, where would you get
the models you can destroy in the interpretation sections of your
observational papers?!?

\vfil

\section{Multi-Frequency Support for High-Energy Observations}

\subsection{Introduction}

In this section I summarize the discussion and conclusions of the
workshop on multi-frequency support of high-energy observation.  I
shall emphasize the importance of long-term monitoring programs, both
in the optical and at radio wavelengths. These are needed not only for
statistical studies but also to get knowledge on the history of
individual sources, the frequency and amplitude of outbursts and so
forth. They are the only appropriate tool to evaluate possible
correlations of outbursts at different frequencies.  Clearly, the
efforts for long-term monitoring in the optical need to be
strengthened.  In contrast, the event-monitoring is regarded as
functioning in principle, but the systems for sharing information 
need improvements.

In some sense now is not the ideal time to consider improvements for
the multi-frequency support of high-energy observations.  Apart from
the OJ-94 project, which was triggered by a theoretical prediction,
the detection of more than 50 AGN by EGRET has been the driving force
for many of the multi-frequency observing campaigns.  The surprising
finding that a subset of AGN, which subsequently has been recognized
to be those objects which are blazars, emit the bulk of their
bolometric luminosity at $\gamma$-ray energies, and that a large
fraction of these objects are variable even on the smallest observable
time scales, has led to the insight that observations in only one
frequency regime are insufficient. The standard $\gamma$-ray
transparency arguments imply that the emission is Doppler-boosted and
thus originates in a jet, similar to the radio synchrotron emission
and part of the optical and X-ray flux. Simultaneously observed
spectra from the radio regime to TeV $\gamma$-rays are therefore the
appropriate tool to discriminate between the models.

However, there is only a little time left to make such observations.
For the next few years the observational coverage at X-rays and above
100 GeV will be very good, but in the range of 1 MeV to 50 GeV there
is no instrument scheduled to be flown past the death of EGRET aboard
the Compton Gamma-ray Observatory.  EGRET is running out of gas, and
besides that is also suffering a slow deterioration in the performance
of the spark chamber. A possible successor named GLAST did well in
NASA project studies, but is still far from funding.  So in the
foreseeable future there will be no instrument which allows us to
continue the multi-frequency spectral monitoring of AGN. We can expect
the \v Cerenkov technique to work at lower energies, possibly down to
30 GeV. These telescopes are more sensitive to small fluxes than
EGRET, but can only observe one source at a time. Furthermore,
absorption by the intergalactic infrared background and possibly
intrinsic cut-offs will prevent a large fraction of AGN from being
observable at or above 100 GeV. Prime targets will be rather  nearby
BL~Lacertae objects as well as galactic transients and superluminal
sources.  The main focus here will be studying rapid variability. A
similar trend is obvious at hard X-rays where most forthcoming
instrument are best suited to, or even dedicated to, timing and
rapid variability observations.

In a few years the field of high-energy observations will experience a
change in direction from spectral monitoring to rapid variability
studies. This implies that the requirements for multi-frequency
support will change accordingly. In these notes I will briefly review
what we have learned technically from recent multi-frequency campaigns
and I will discuss ways to improve the support of the remaining EGRET
observations of blazars.

\subsection{Requirements for multi-frequency monitoring}

This discussion is split into two parts, the first is devoted to
long-term monitoring and the latter deals with coordinating
observations in response to an unusual event of any kind.

\subsubsection{Long-term monitoring}

Long-term monitoring data on individual sources are extremely
important for determining the degree of correlation between
high-energy outbursts and emission at other wavelengths.  Not only is
knowledge of the flux level at outburst and in quiescence necessary,
but also a determination of the frequency of outbursts and the source
variability time scales. It is my personal view that in some of the
early claims of correlated variability between $\gamma$-rays and the
optical or radio emission we have been too uncritical. We neglected the
distinction between correlated activity in the general sense, which we
often find, and correlations of rapid events. Some AGN have high
optical fluxes when they are also bright at $\gamma$-rays. But does
that allow us to relate the one-day $\gamma$-ray flare to the one-day
optical flare we observed a few days later? This is an issue beyond
the inaccuracy of the term correlation.  It is rather an illustration
of the fact that a careful consideration of the multi-frequency
history of a source is required, which can only be done on the basis
of long-term monitoring data. Finally there is no question that such
data are also valuable for statistical studies.

Everyone who has experience with program committees knows that the
best and only way to perform monitoring of many sources over a long
time is to have no such committee involved. Dedicated instruments seem
to be available in the radio regime with the Metsah\"ovi station and
the telescope of the University of Michigan. The situation in the
optical is less advantageous.  This leads to situations like that of the
OJ-94 project where a major fraction of the optical data was provided
by Paul Boltwood, i.e. by an amateur who observes with a 7-inch
telescope and a self-made CCD in the backyard of his house in
Ottawa. As a result  optical long-term coverage for most sources 
is insufficient. One way to go may be small robotic telescopes which
need limited manpower to operate. Another way could be to get
Colleges and more amateurs involved. A problem with the latter
strategy surely will be the required effort for the selection of
sufficiently serious and careful amateurs.  Also, scientists have
expressed their interest in optical polarization data.  It is doubtful
whether such difficult measurements can be reliably done by amateurs.

The most important issue for the moment however is a harmonization of
target lists. It seems to me that currently each observatory has its own
list of monitoring sources with little or no overlap between the lists
at different frequency regimes. As far as support for high-energy
observations is concerned this is not a satisfying situation. We
strongly recommend that all groups who do monitoring campaigns,
especially in the optical, include a set of `common' sources in their
target list, e.g. the AGN detected by EGRET or the compact sources
observed by \v Cerenkov telescopes. This still allows each group to do
their research programs on well-defined samples, but on the other hand
makes sure that for a group of interesting sources good long-term
light curves in the optical and in the radio range are produced.

\subsubsection{Event monitoring}

With the advent of EGRET and its detection of many AGN as strong and
variable emitters of high-energy $\gamma$-rays, observers throughout
the world have become interested in organized multi-frequency
campaigns that parallel high-energy observations. Unfortunately, the
sources have often been ``on vacation'' during the observations,
showing little or no activity, so that Target-of-Opportunity
observations have become more en vogue. In this mode unusual activity
at high energies trigger closely sampled observations in the optical
and radio range or vice versa. Correlated observations at X-rays and
$\gamma$-rays still have to be organized in advance and are undertaken
somewhat reluctantly by the observatory program and time allocation
committees, with the exception of a few highly promising sources.

The event monitoring has brought some interesting results though in
many cases the studies suffer from missing long-term light
curves. Since in the most extreme case an outburst has been observed
on a time scale of half an hour, one will want to obtain a very close
sampling in the multi-frequency follow-up observations with about
sub-day time scale in the optical and day time scale at radio
frequencies. This implies that many observatories, especially in the
optical, have to be involved.

Communication clearly is the key in this business, and that in two
senses.  The first sense is that a functioning alert system is
required, which spreads the information on an unusual event to
observatories throughout the world, similar to the Bacodine system for
bursts. In the OJ-94 project such a system has proved its
capabilities with observers who had agreed before to serve on
this campaign. In the general case the effect of surprise will be much
larger. What we need in future is a system which can be triggered by a
number of experiments from X-rays to TeV $\gamma$-rays, and which is
faster than the IAU telegram system, since we will not want to wait
for two days before we start multi-frequency observations for a
one-hour TeV flare. This will be especially true when EGRET is no more
and we observe nearby BL Lacs and Galactic sources at hard X-rays and
TeV $\gamma$-rays.

The second sense in which communication is important concerns the
availability of observing schedules for high-energy observatories. A
large fraction of the observatories make their schedules publicly
available somewhere on the Web, but there is no mother page which
provides links to all the observatory pages.  As a result the system
is somewhat unorganized. Public schedules could also give some advance
warning that an outburst may be observed, and thus help observers in
monitoring the event.  The problem with such a mother page definitely
is that someone has to maintain it. Each one of us felt the necessity
of an organizer on the Web, but no one volunteered to do the work.

\subsection{Conclusions}

In this section I have discussed a few problems and their possible
solution for the multi-frequency support of high-energy
observations. We have seen that information is the {\it conditio sine
qua non} in this business. Everyone in our discussion panel expressed
their commitment to improving the communications networks and fast
response capabilities of the community as well as continuing and
expanding long-term monitoring programs.  The two main multi-frequency
campaigns in the remainder of the EGRET era (10 weeks on 3C279 and 4
weeks on 0528+134) will be a test of how efficient we can get in the
near future.

With the already discussed change of direction in high-energy
observations fast approaching, the requirements for the
multi-frequency support can only get more demanding. Short time scale
phenomena need closely sampled light curves to be studied.  On the
other hand, for Galactic sources long-term monitoring will become less
important, partly since the sources are often absent between outbursts
and partly since outbursts are generally temporally well
separated. This is advantageous in the light of the low typical radio
fluxes in the milli-Jansky range which require synthesis telescopes
and hence committees.  The interests of the long-term monitoring
people and the interests of the event-monitoring people will diverge
slightly, but that doesn't imply that the requirements for the
monitoring get less difficult. Galactic sources tend to be active for
several weeks with a lot of substructure in the light curves. The most
prominent example is surely the 470 keV feature of Nova Muscae 1991,
which was observed to last for only about ten hours. Here event
monitoring will need sampling on sub-day time scales sustained over
several weeks.  This requires a high availability of telescopes,
especially at optical wavelengths.  This together with the short
advance warning time for outbursts will probably favor robotic
telescopes once they can be produced in series at a sufficiently high
level of mechanical and electronical stability.

Finally let me spend a few lines on the interaction with the rest of
scientific community, with special emphasis on theorists. What we have
done so far is mainly provide simultaneously observed multi-frequency
spectra of objects at a given time, which then can be compared to what
the models predict. To no one's surprise, all models have been able to
fit these spectra, irrespective of whether the source was in the state
of an outburst or in quiescence.  The drawback here is that we have
often an equal number of model parameters and true degrees of freedom
in the data. We have seldom seen model fits which used the same basic
parameters such as Lorentz factor, inclination angle, accretion disk
luminosity, and so forth to reproduce the multi-frequency spectra of a
source at various states. What we require in the future is
multi-frequency spectra at different stages of an outburst as well as
the theory predictions for these. I would like to see how an
instantaneous release of energetic particles would find its way
through the electromagnetic spectrum in the Inverse-Compton models and
the Proton-induced-cascades and compare that to what we see during
outbursts. Theorists have just started to work on such calculations,
and I am sure that in a few years this kind of research will prove to
be highly valuable for the understanding of the blazar phenomenon.

\section{Monitoring of Blazars}

\subsection{LONG-TERM SYNOPTIC MONITORING} 

The ``monitoring'' group discussed the question of whether long-term
optical monitoring might be approaching the point of diminishing
returns. The Florida team, for example, now has 27 years of internally
consistent data on as many as 200 sources. What will we learn from
extending observations of this kind? While there was no general
consensus, I pointed out that a number of the Florida objects
display long-term, possibly cyclic, changes in their base levels that
are not defined even by a quarter-century of data. It was further
observed that monitoring of this kind frequently detects outbursts
that then become the subject of intensive and highly profitable
campaigns by the entire AGN community.

Several participants argued for the establishment of a dedicated
global network of moderate-size telescopes for the purpose of
providing 24-hour coverage of a limited number of objects, especially
during outburst phases. There was discussion of whether such
instruments should be manned or robotic, although the cost and
complexity of the latter was recognized. There was a general feeling
that the interested members of the community should undertake to
organize and lobby for funding for the establishment of such a
network. I suggested that such a network already exists in the
form of numerous underutilized working telescopes, with willing
observers who need only to be organized, encouraged, and provided
with very nominal funding. A question was raised as to whether those
engaged in monitoring should agree on a single frequency band unless
they undertake multi-frequency observations. While this did not
provoke extended discussion, one point of view was that the $V$~ band
would be preferable for historical reasons, such as tying in with
earlier photographic observations; however, it was pointed out that,
given the characteristics of the now universally-employed CCD
detectors, the $R$~ band is easier to use because of the much shorter
exposures it allows. Meg Urry has argued for the establishment of a
dedicated space telescope for 24-hour optical observations of AGN;
there was a rather general feeling that the costs of constructing,
launching, and maintaining such an observatory would dwarf the cost of
the proposed ground-based network.

It was pointed out that with the demise of the International
Ultraviolet Explorer spacecraft, and the incipient demise of the EUVE
spacecraft, the AGN community is losing far-ultraviolet coverage of
its sources. However, a counter-argument was put forward that, since
it has been shown that there is a close correlation between the far-uv
and visible-band observations, the UV lacuna can be filled by better
ground-based observations.

Several times during the Workshop the question was raised as to
whether one might rely on amateur astronomers to augment the
professional ranks in providing AGN data for either long-term
observations, or during short-term campaigns. A shining example is
the beautiful observations provided by the Canadian amateur Paul
Boltwood, who presented his work during the conference and was present
at this workshop. Paul noted that he had spent tens of thousands of
dollars and thousands of hours constructing, automating, and operating
the Boltwood Observatory. It was his educated opinion that extremely
few amateurs would be in a position, either time-wise or dollar-wise,
to duplicate his effort. He is, further, a professional computer
expert, which played a major role in automating the Boltwood
Observatory.  Paul, however, very kindly volunteered to advise any
other advanced amateurs who might wish to follow in his footsteps.

\subsection{MULTI-FREQUENCY CAMPAIGNS}  

As in the case of long-term monitoring, the question was raised as to
whether we are reaching the point of diminishing returns with
intensive multi-frequency campaigns centered on outbursts of active
sources. Dick Miller expressed the opinion that we are indeed reaching
such a point of saturation unless - and this was an important
qualification - new techniques are brought to bear during the
campaigns. There was general agreement that multi-frequency campaigns
would be of much greater value if intensive monitoring were carried on
for several months surrounding an outburst.  This would aid in placing
the flare in context in the midst of more normal behavior. It was of
course recognized that initiating a campaign in advance of an outburst
would, in our present state of understanding, require the use of a
crystal ball!

\subsection{MICROVARIABILITY} 

It was recognized that the present popularity of microvariability
studies is at least in part an artifact of the realities of telescope
scheduling. It is much more feasible to obtain a few consecutive
nights during which microvariability runs of a few hours each can be
conducted, than to obtain the many nights scattered over a long period
of time that are required for long-term studies. Dick Miller pointed
out that thus far there have been no real microvariability
``campaigns.''  He urged the desirability of such campaigns that would
provide 24-hour continuous coverage, without gaps, for periods of at
least a few days. Needless to say, this would require the
participation of observatories appropriately spaced in longitude, with
sufficient redundancy to mitigate the effects of local weather. Based
on his own extensive experience in the field, Dick suggested that
definitive and unambiguous microvariability results require observers
to aim for errors in the range of 0.01 to 0.02 magnitude, with an
upper limit of 0.04 magnitude. The exposures required to achieve such
precision then of course place constraints on the time resolution that
can be expected during a run.

\subsection{Database}  

The group discussed at length the desirability of establishing a
common database for pooling information on AGN variability.  There
was general agreement that this would be a valuable resource, but it
was also recognized that there are a number of problems with
implementing it. A central problem, of course, is that of protecting
the rights of the original observer until he/she has obtained adequate
credit; this is especially crucial where the work is funded by an
agency that expects to receive due recognition for its investment. If
such problems can be solved, an attractive possibility is to collect
the observations in a site on the World-Wide Web, which can be
accessed immediately by all users. An additional difficulty is the
question of format in presenting the data. An extended discussion
concluded that if the burden of converting to a common format were
placed on the observers, this added work would have the undesirable
effect of discouraging participation. The final recommendation was
that observers should contribute their data in their own formats, with
any necessary conversion s being left up to those who wish to use
it. A suggestion was made that instead of tabulating the observations,
the web site should merely list the names and addresses of observers
in a position to supply the observational results; this alternative
proposal did not receive strong support.

\section{Host Galaxies and Environments}

\subsection{Introduction}

While the other workshops were focusing on direct manifestations of
blazar activity (jets and the emission at all wavelengths that comes
from the jet and whatever is the ``central engine'' of blazars), a
small group of us were charged to discuss the ``houses'',
``neighborhoods'', and ``hometowns'' of blazars.  The motivations for
studying the environments of blazars (which in theory we took to
include BL Lac objects, OVVs, and variable high polarization quasars,
but in practice we tended to focus on BL Lacs) are to identify the
parent populations of a phenomenon that is believed to be strongly
aspect-dependent, to understand whether or not certain environments
are necessary for the ``birth and maintenance'' of a blazar, and to
investigate any effect the blazar has on its home (e.g. triggering
star formation, etc.). Perhaps just a reflection that the previous
three days of talks and posters had already covered our current
understanding of blazar environments, our workshop discussions only
briefly touched on the current state of knowledge (summarized in the
next subsection) and instead dwelled on what questions we would like
to address with future observational programs.  Here we were not shy,
concluding that we would like to know more about the environments in
all dimensions -- physical (1 kpc to 1 Mpc) and temporal (from the current
epoch to high redshift).  Some of the many questions we considered and
examples of the possible observational programs that might provide the
information we desired are discussed in the third subsection of this
summary.

\subsection{What We Think We Know}

There has been a great deal of progress made in just the past few
years in efforts to image the host galaxies and cluster environments
of BL Lac objects from the ground (see various contributions to these
proceedings, as well as Falomo 1996, Wurtz et al. 1996 and
included references). The {\it Hubble Space Telescope} is also
beginning to contribute to the study of the host galaxies (see
contributions to these proceedings by Treves and Jannuzi and
Falomo et al. 1997 for examples).  While the sample of objects that have
been well observed remains small (well under 100), the data are
generally 
consistent with the host galaxies of BL Lacs being elliptical or bulge
dominated galaxies that are similar, perhaps identical, to those of
FR~I radio galaxies. There is still debate over whether all BL Lacs are in
ellipticals and whether the distribution of host galaxy properties is
more similar to the host galaxies of FR~I or FR~II radio
sources. However, convincing spiral hosts have proved extremely rare
(only two current examples, see for discussion Wurtz et al. 1996) and the
proper classification of these two objects is still being debated.  The
imaging programs are also yielding information on the incidence of
close companions, allowing the first investigations of the role of
interactions and mergers in triggering and feeding blazar activity,
but this work is just beginning.

Progress is also being made in studying the larger scale environments
through wide-field imaging of BL Lacs and control fields.  Again, the
samples being studied, while small, are increasing in size and appear
to confirm that BL Lacs avoid rich clusters, but might still be drawn
from a subset of FR~I radio sources (see e.g., Wurtz et al. 1996).

\subsection{What More Do We Want To Know?}

While the progress made in imaging studies of the hosts and
environments of BL Lacs has been significant, there are possible
problems with what has been done and a great deal more that needs to
be done. This is perhaps best demonstrated by listing some of the many
questions that came up during our discussions:

Given the experience from comparing HST and ground-based observations
of quasar host galaxies, how robust are the existing results and are
additional checks warranted? For discussion of the comparison of
ground-based and HST results see for example the work of Bahcall
et~al. (1997) and references therein, and McLeod and Rieke (1995) on
the host galaxies of AGNs. In particular the discussions on the
problems of extracting useful measurements on the hosts -- problems
that only increase when the angular scales of the objects are smaller,
as is the case for more distant objects -- when the nucleus is very
luminous.

Even with excellent imaging results, how do we know that all of the
surrounding nebulosity we see around blazars is star light? Is any of
radiation scattered light or emission lines from gas? If the
surrounding nebulosity is produced by stars, are they old, young, or a
mix of populations? Are the companion objects seen in some images
really close companions or substructure in the host galaxy? If
galaxy--galaxy interactions are really critical in triggering a
blazar, how strong an interaction is needed and what implications does
this have for the birth rate as a function of cosmic time and the
lifetime of individual objects?

Do all blazars have broad line regions (BLRs)? If they do, what is the
range of the properties of the blazar BLRs and how do these properties
compare with those of the possible parent populations?  If they do not,
why not?  Is there a very hot gas component in/around blazars?  Are
blazars at the center of cooling flows?  How does fuel get to the
central engine?

Are all blazars members of a single class (requiring that BL Lacs,
OVVs and variable highly polarized quasars all come from the same
parent population, which seemed unlikely to those of us at the
workshop) or do the environments of these different subclasses provide
clues to why the objects appear to have some significant differences
(e.g. see work of Gabuzda and collaborators for some possible
differences), despite sharing the dramatic blazar characteristics?

Do the properties of blazars at high redshifts ($z>0.6$) differ from
those of low-redshift samples?  Do significant numbers of
high-redshift blazars even exist?  Can BL Lacs and blazars be related
to other AGNs in an evolutionary- or luminosity-based sequence?

\subsection{How Do We Get the Answers?}

The end of the 1990's and the next century will see continued growth
in the observational capabilities available to the astronomical
community and enable many of the questions listed above to be
answered.  In summarizing the ideas from our workshop for the entire
conference,  we decided to provide examples of how new observations and
observing capabilities could be brought to bear on some of the
unanswered questions, rather than trying to provide guides to
answering all of the questions we had listed.  We tried to highlight
in each case how the proposed observations could lead to the
determination of physical properties of the blazars or their
environment. We have reproduced three of the examples below.

\subsubsection{Spectroscopy With Large Telescopes}

Extensive spectroscopic follow-up of the light from blazar host
galaxies, close companions, and host cluster members will provide
answers to many of the unanswered questions. Lots of photons (i.e.,
big telescopes) and high spatial resolution (excellent site,
telescope, and instruments) are necessary for success.  When all of
these things are available, as is clearly demonstrated by the
contribution from Joe Miller to this conference with his observations
from the Keck~I telescope, one can easily obtain excellent spatially
resolved spectra of the {\it star light and H~II regions} in the host
galaxies of quasars. It should only require the increased availability
of large-telescope observing time (which should be more widely
available once the VLT, Gemini, and other telescopes are in
operation), a few proposals, kind comments from time allocation
committees, and some good weather before large numbers of blazar host
galaxies can be observed in a similar manner. Studies of the stellar
populations of the host galaxies and investigations of the role of
interactions in blazar activity should therefore make rapid progress
in the near future.

\subsubsection{Observations at Non-Optical Wavelengths}

The masses of the host clusters as well as the presence of hot gas and
``warm absorbers'' in BL Lacs can be studied with X-ray telescopes,
and we expect continued progress in these investigations using both
existing and future (e.g., AXAF) X-ray observatories.  Information on
smaller spatial scales should continue to become available from
observations at millimeter through radio wavelengths, including
studies of Faraday rotation as a diagnostic for hot plasma surrounding
the central source. Millimeter, sub-millimeter, and IR observations
will provide information on the dust and gas content of the host
galaxies. Maps of the gas and dust content of the hosts, when combined
with high spatial resolution imaging and spectroscopy of the stellar
light (which will eventually be aided through the use of coronographs
and/or HST) might allow detailed investigation of the structure of the
host galaxies and studies of the available fuel reservoirs of the
blazars.

\subsubsection{Extensive New Surveys}

In addition to discussing how to answer the questions listed in the
previous section, we also discussed how our current knowledge of the
environments of BL Lacs might be biased and how this might be
corrected with additional work.  First, most of our knowledge of the
host galaxies and cluster environments of BL Lacs comes from studies
of low redshift ($<0.6$) objects and those BL Lacs which do not have
large ratios of observed nuclear brightness to surrounding nebulosity.
This is not surprising, and may simply be a consequence of studying
biased samples. Existing samples are flux limited, and are therefore
biased against finding all but the most luminous distant objects. Even
the best existing blazar host galaxy studies have relied on samples
whose primary selection criteria depended on blazar properties
(selection at wavelengths where the radiation is dominated by the
non-thermal beamed emission), and are therefore further biased towards
the extreme members of the population. Finally, our ability to detect
the surrounding nebulosity requires very high spatial resolution and
extremely good dynamic range.  A partial solution, although expensive
in terms of effort and telescope time, is to construct even larger
samples of blazars by extending follow-up of x-ray and radio surveys
to even lower flux levels. This would allow the construction of large
samples with a larger dynamic range in the source properties.  An
obvious candidate for follow-up is the FIRST radio survey, currently
being carried out using the VLA, but there are certainly other
suitable surveys in progress or planned.

\subsection{Exciting Times}

In general the workshop participants were excited about the prospects
for significant progress in the future. The most significant challenge
to progress that we identified was not scientific, but rather the need
to effectively communicate to the telescope time allocation committees
and the research funding agencies the nature and degree of progress
that is being made in the study of blazars so that this work will be
enabled in the future.  This later discussion, as was appropriate, took
place over some good wine and beer, rather than during the main
session of the workshop.

\section{A Very Well Deserved Thank You!}

All of the workshop chairs would like to thank, on behalf of all the
conference participants, the wonderful organizers of the conference.
The meeting was informative, productive, and enjoyable only because of
the extensive preparation and extremely hard work of both the
scientific and local organizers.  Special thanks have to be given to
Jose Antonio de Diego, Sumpsi Montagut, and the main driver for the
entire meeting, Mark Kidger. Thank you!

\end{document}